\documentstyle[12pt]{article}
\title{CRITICAL BEHAVIOR OF THE ANTIFERROMAGNETIC HEISENBERG MODEL ON
A STACKED TRIANGULAR LATTICE}
\author{T. Bhattacharya\thanks{Present address:
MS B285, Group T-8, Los Alamos National Laboratory, \ \
NM 87544},
A. Billoire, R. Lacaze{\raisebox{2pt}\dag}
and Th. Jolic{\oe}ur\thanks{Chercheur CNRS}\\
Service de Physique Th\'eorique de Saclay\thanks{
Laboratoire de la Direction des Sciences de la Mati\`ere du CEA.}}
\date{September 24, 1993}
\begin{document}
\maketitle
We estimate, using  a large-scale Monte Carlo simulation, the critical
exponents of the antiferromagnetic Heisenberg model on a stacked
triangular lattice. We obtain the following estimates:
 $\gamma/\nu= 2.011 \pm .014 $,
$\nu= .585 \pm .009 $. These results contradict a perturbative
 $2+\epsilon$ Renormalization Group calculation that points to
Wilson-Fisher O(4) behaviour.
While these results may be coherent with $4-\epsilon$ results from
Landau-Ginzburg analysis, they show the existence of an unexpectedly
rich structure of the Renormalization Group flow as a function of
the dimensionality and the number of components of the order parameter.

\vskip 2.cm
\rightline {SPhT-93/015}
\rightline {cond-mat/9302019}
\rightline {Submitted to: Europhysics Letters}
\vfill\eject

\section{INTRODUCTION}

There is  at  present a satisfactory understanding of the critical
behaviour of physical systems where the rotation symmetry group
O(N) is broken down to O(N-1) at low temperatures.
Several theoretical tools are available to estimate the critical
exponents and there is good agreement between these estimates.
The Wilson-Fisher fixed point which describes the critical physics can be
smoothly followed between two and four dimensions: for $N\ge 3$
the $2+\epsilon$ and $4-\epsilon$ renormalization group
expansion merge in a continuous manner.

The situation is much more
complicated  when the rotation symmetry is {\it fully} broken in the
low-temperature phase. A prominent example is found in the so-called
helimagnetic systems where Heisenberg spins are in a spiral arrangement
below the critical temperature. It is an interesting question, both
theoretically and experimentally, to know the  corresponding universality
class.
A widely studied prototypical model is the antiferromagnetic Heisenberg model
on a stacked triangular lattice, which is simple and
display commensurate helimagnetic order below a transition point $T_c$.
There is little consensus in the literature
regarding critical phenomena associated with this model
\cite{review}.

This topic has been investigated by use of a $D=4-\epsilon$ renormalization
group calculation\cite{4d}. The corresponding Ginzburg-Landau theory for a
N-component vector model involves two N-component bosonic fields. It is found
that for N large enough the transition is second order and not governed by the
Heisenberg  O(2N) Wilson-Fisher fixed point but by a different fixed point
which is also non-trivial for $D<4$.
For smaller N,  this new fixed point disappears and there is no stable
fixed point which is an  indication for
a fluctuation-induced
first-order transition. The dividing universal line between second-order and
first-order behaviour is found to be $N_c (D)= 21.8-23.4\epsilon
+O(\epsilon^2)$. The rapid variation of $N_c$ leaves us rather uncertain
about the fate of the case $D=3$.
Clearly more information is needed about the $N_c (D)$ line in the
(number of components-dimension)-plane.

A $D=2+\epsilon$ renormalization group study has been performed
for a system of Heisenberg spins\cite{2d} by use of a non-linear sigma model
defined on a homogeneous non-symmetric coset space $O(3)\times O(2)/O(2)$.
It was found that near two dimensions the system undergoes a second
order transition which is governed by the O(4) usual Wilson-Fisher fixed point.
In fact the symmetry $O(3)\times O(2)/O(2)$ is dynamically
enlarged at the critical point to $O(3)\times O(3)/O(3)$ and
$O(3)\times O(3)$ is O(4).
This mismatch between the expansions near four dimensions and near two
dimensions is quite unusual and does not happen in the well-studied
$O(N)\to O(N-1)$ critical phenomena. As a consequence,
the $D=3$ case remains elusive
and a direct study in three dimensions is called for.

Some preliminary Monte Carlo (MC) simulations
have shown  evidence \cite{kawamura1,kawamura2,kawamura3,kawamura4}
for a continuous transition in the case of
the Heisenberg antiferromagnet on a stacked triangular lattice.
The exponents found are not compatible with those of the
O(4) vector model in three dimensions.
We did a large-scale simulation with much better statistics
than previous attempts. As a consequence we are able to pin down
the transition temperature in a very precise manner and to
obtain reliable estimates of critical exponents.

We thus focus on the classical spin model defined by the
classical Heisenberg Hamiltonian:

\begin{equation}
H= \sum_{<ij>} J_{ij} \   {\bf S}_i \cdot {\bf S}_j .
\end{equation}

The exchange interaction $J_{ij}$ is nonzero
({ $J_{ij}=1$  in what follows}) between nearest-neighbors of a
stacked triangular  lattice and the spins are three-component unit vectors.
The classical ground state is found by minimizing the Fourier transform
$J({\bf Q})$ of the couplings $J_{ij}$. The spins adopt a planar arrangement
on a three-sublattice structure with relative angles 120 degrees.

\section{The simulation}

We made two successive sets of Monte Carlo simulations of this model,
using  slightly asymmetric lattices
(with periodic boundary conditions) of shape
$L^2 L_z$, where $L$ is the linear size inside the planes, and $L_z=2/3 L$
the stacking size.

The first set of simulations was run on a CM-2 8K massively parallel computer.
We made runs on lattices of size $12^2 \ 8$ (with a total of
600 000 Monte Carlo sweeps of the lattice), $24^2 \ 16$ (with a
total of 1 500 000 sweeps) and $48^2 \ 32$ (with a total of 1 570 000 sweeps).
In the first case we used the Heat-Bath algorithm, in the later two cases
we used an  Hybrid Overrelaxation algorithm\cite{adk} where each Heat-Bath
sweep  is followed by an energy conserving sweep.
The next set of  simulations was run on CRAY vector computers
on lattices of size
$18^2 12,\ 24^2 16,\  30^2 20 \ , 36^2 24$ and $48^2 32$ with 4 to 8  $10^6$
Hybrid Overrelaxation sweeps.
Both simulations concentrate in the immediate vicinity of the transition.

Let us note by  ${\bf S}_a$ the total  spin per site for sublattice
$a$ ($a \in [1,2,3]$),  we measured the magnetization:

\begin{equation}
M_L(\beta)= {1\over 3} \sum_{a} <|{\bf S}_a|> ,
\end{equation}

the susceptibility:

\begin{equation}
\chi_L(\beta) = {L^2 L_z\over 3}\ \ ( \sum_{a} <{\bf S}_a^2>-<|{\bf S}_a|>^2 ),
\end{equation}

and the fourth-order cumulant:

\begin{equation}
B_L(\beta)=1 - {1\over3} \sum_{a}  {<{\bf S}_a^4> \over <{\bf S}_a^2>^2} .
\end{equation}

Our strategy to extract the critical exponents was first to
estimate the value of the critical temperature $\beta_c$,
as the point where $B_L(\beta)$ is $L$ independent and then to
estimate  the exponents from the following set of equations:

\begin{eqnarray}
\chi_L(\beta_c) &\sim& L^{\gamma/\nu} ,\\
M_L(\beta_c) &\sim& L^{-\beta/\nu}     ,\\
{{\partial B_L(\beta)} \over {\partial \beta}}\Biggr|_{\beta=\beta_c}
&\sim& L^{1/\nu} \\
{\partial \over {\partial \beta}}\ln M_L(\beta)\Biggr|_{\beta=\beta_c}
&\sim& L^{1/\nu}.
\end{eqnarray}

The extrapolation, from the $\beta$ value used for the simulation,
to $\beta_c$ is done using the Ferrenberg-Swendsen
technique \cite{Ferrenberg_Swendsen}. This technique is invaluable
to extrapolate in a neighborhood of size $ \sim 1/L^{1/\nu}$
(We understand that, when used blindly outside
such a tight range, it may give wrong results).
An alternative strategy would be to evaluate exponents from the
finite size scaling of the maxima of $\chi_L(\beta)$, $M_L(\beta)$,
${{\partial}\over {\partial \beta}} B_L(\beta) $
and ${\partial \over {\partial \beta}}\ln M_L(\beta)$.
It would however require to perform
simulations at four $\beta$ values, close to the four ($L$ dependent)
points where this quantities reach their maxima,
otherwise accuracy would be lost.
The statistical analysis is done with respect to 20 bins, using
first-order bias-corrected jackknife (see e.g. Ref.\cite{Paper_q10} and
references therein). The first 20~\% of each run is discarded
for thermalization.

{}From simulations of lattices of increasing sizes $L_1 < L_2 <
\dots$ converging estimators of $\beta_c$ are obtained by solving
the equations:

\begin{equation}
B_{L_{i-1}} (\tilde\beta) = B_{L_{i}} (\tilde\beta) .
\end{equation}

\begin{table} [tbh]
\begin{footnotesize}
\begin{center}\begin{tabular}{|c|c|c|c|c|c|}\hline
  & $\beta_c$ & $\gamma/\nu$ & $D-2\beta/\nu$ & $\nu$ & $\nu$ \\
[3pt]\hline
24-12  & 1.04401 (38) & 2.009 (14) (04) & 2.030 (07) (07) & .651 (22) (03)
							  &  .604 (10) (01) \\
48-24  & 1.04462 (21) & 2.033 (18) (14) & 1.998 (09) (25) & .590 (20) (04)
							  & .576 (10) (04) \\
[3pt]\hline\hline
24-18  & 1.04452 (40) & 2.010  (26) (06)  & 2.011 (08) (10) & .605 (30) (03)
							    &  .608 (17) (02) \\
30-24  & 1.04388 (36) & 1.994  (36) (11)  & 2.021 (13) (16) & .578 (35) (03)
							    & .583 (19) (03) \\
36-30  & 1.04408 (37) & 1.988  (55) (16)  & 2.020 (18) (21) & .605 (55) (12)
							    & .607 (30) (04) \\
42-36  & 1.04485 (25) & 2.083  (68) (14)  & 1.984 (20) (31) & .528 (48) (06)
							    & .548 (28) (06) \\
48-42  & 1.04427 (30) & 1.986  (61) (18)  & 2.004 (22) (39) & .639 (69) (05)
							    & .577 (27) (03) \\
[3pt]\hline
48-18  & 1.04432 (06) & 2.011  (07) (12)  & 2.010 (03) (21) & .590 (09) (06)
							    & .588 (04) (03) \\
[3pt]\hline
all    &    N/A       & 2.011 (07) (12)  & 2.011 (02) (19)  & .585 (07) (06)
							    & .585 (04) (04) \\
\hline
\end{tabular}\end{center}
\end{footnotesize}
\label{CRAY}
\caption{Critical temperature and critical exponents estimates, using
equations 9, 5, 6, 7 and 8 respectively.
First two lines are CM-2 data, next lines the CRAY data.
The first column gives the linear sizes  of the  lattices
used. The two numbers inside parenthesis are
the estimated statistical errors, direct and induced,  on the last
 digits of the number on their left.}
\end{table}

The results can be found in the second column of Tab.1.
Within our statistical accuracy all estimates are compatible, there is
neither any clear lattice size dependence, nor any
discrepancy between the two sets of runs. Our final estimate is
(the quoted error is deliberately on the safe side)
\begin{equation}
\beta_c =1.0443 \pm .0002 .
\label{betac}
\end{equation}

We use this value to  compute $\chi_L(\beta_c) $, $M_L(\beta_c)$,
${{\partial} \over {\partial \beta}} B_L(\beta)$
and ${\partial \over {\partial \beta}}\ln M_L(\beta)$ in
order to estimate $\gamma/\nu$,  $\beta/\nu$ and $\nu$ .
Results can be found  in Tab.1.
We quote separately the  estimated ``direct'' statistical errors computed
from the dispersion of the results from the 20 bins, and the
errors induced by the
uncertainty on the determination of $\beta_c$, computed
as the (absolute value of the) difference
between the values obtained using our best estimate of
$\beta_c$ and the values  using the
one standard deviation estimate from Eq.\ref{betac}.
The last line in Tab.1 gives the results of  linear fits of the CRAY data for
$\ln(\chi_L(\beta_c)) $, $\ln(M_L(\beta_c))$,
$\ln({{\partial } \over {\partial \beta}}B_L(\beta)\biggr|_{\beta=\beta_c})$
and
$\ln( {\partial \over {\partial \beta}}\ln M_L(\beta)\biggr|_{\beta=\beta_c})$
respectively. The results of the fits are stable against omitting
the smallest  lattice data. The $\chi^2$ is equal to 1.5, 3.7, 1.9
and 4.5 respectively (we expect $\chi^2 \approx 4$).
Such values mean that linear fits give good representation of
the data for $L \geq 18$ (with the current accuracy).

Our final numbers  are
$\gamma/\nu= 2.011 \pm .014 $,  $D-2\beta/\nu=2.011 \pm .019$,
$\nu= .585 \pm .009 $.
We observe that hyperscaling is verified within errors,
and $\eta$ is very small ($|\eta| \approx .01$).
The value for $\nu$ is clearly not compatible with the value for
the $O(4)$ fixed point\cite{LG} ($\nu \sim .75$).
%We have thus obtained
%excellent evidence for a continuous transition in the $D=3$ case and we
%have shown that the critical exponents are not compatible with
%the RG prediction of the $2+\epsilon$ expansion.

Published Monte Carlo results for $\nu$ are $.53 \pm .02$\cite{kawamura1},
$.53 \pm .03$\cite{kawamura2}, $.55 \pm .03$\cite{kawamura3} and
$.59 \pm .02$\cite{kawamura4}.
Our results are in good agreement with the most recent results
(and with the result $.57 \pm .02$ from a model of commensurate
Heisenberg helimagnet\cite{diep}).
The methods of analysis are quite different. Ref.\cite{kawamura4} uses
data taken in a wide region around the  transition point
and adjust the values of the
critical exponents using the old fashioned ``data collapsing''
method. This method has the disadvantage to give much weight to
points with large value of $(\beta-\beta_c) L^{1/\nu}$. We extract the
exponents directly from data reweighted to $\beta_c$.
We have also a much higher statistics: we use $8\times 10^6$ heat-bath
+ energy-conserving sweeps whereas \cite{kawamura4} uses 6-20 times 20000
sweeps
of a less efficient algorithm. Our estimated statistical error on
the determination of $\beta_c$ is
one order of magnitude smaller than the one in ref.\cite{kawamura4}.

\section{Conclusion}
Let us first summarize the knowledge obtained from perturbative RG studies
as a function of the dimensionality and the number of components of
the order parameter.

\noindent
i) There is a universal line $N_c (D)$ separating a
first-order region from a second-order region near $D=4-\epsilon$.
Its slope is known from RG studies\cite{4d} as well as the critical
value $N_c$ for $D=4$.

\noindent
ii) Large-N studies\cite{large_n} indicate
that the stable
fixed point found above $N_c (D)$ persists smoothly in the region $N=\infty$
and $2 < D < 4$.

\noindent
iii) Smoothness along the $D=2$ vertical
axis can be shown by studying the nonlinear sigma model suited
to the N-vector model: it is built on the homogeneous space
$O(N)\times O(2)/O(N-2)\times O(2)$\cite{Delduc}.
One finds a single stable fixed point
for {\it all} values of $N\ge 3$. In the large-N limit the
exponents from this sigma
model are the same as those of the linear model.

If we believe in the perturbative RG results, then necessarily
the universal line $N_c (D)$ can only intersect the horizontal axis
$N=3$ between D=2 and D=4. The simplest hypothesis is then that
the plane (N, D) is divided in two regions by the line $N_c (D)$.
This line may intersect the $N=3$
axis at a critical dimension $D_c$.
Our Monte-Carlo results thus
imply that $D_c$ is between three and four dimensions since
we observe a continuous transition at D=3.
A possibility would be that the critical behaviour we observe is
described by the fixed point found in the neighborhood of D=4.

However one has to note that the sigma model approach shows
that this fixed point is O(4) for N=3. This is clearly ruled
out by our data. The simplest scenario suggested by i-iii above
is wrong.
It may be that the whole sigma model
approach breaks down since perturbative
treatment of a sigma model neglects global aspects. Previous
studies indeed have performed, as usual, only perturbation theory
for spin-wave excitations\cite{2d}. One may invoke the peculiar vortices
present in the model\cite{vortex}
 since $\Pi_1 (SO(3))=Z_2$ that form loops in
D=3 as responsible of the breakdown of the sigma model. However
vortices are also excluded from the perturbative calculations near D=4
and thus it would be difficult in this case to believe in what is
found perturbatively.
It may be also that the fixed point found in $2+\epsilon$ expansion is
destabilized by some operators containing high powers of gradients
\cite{Chakra}.

We have thus obtained the critical behaviour
of a magnet with a canted ground-state.
This calls for a deeper understanding of the Renormalization Group
behaviour of canted systems since there is no obvious relationship
between the different perturbative approaches,
contrary to the case of collinear vector magnets.
We note for the future that
it would be interesting to obtain the critical behaviour of the XY
canted systems, much more relevant to the experimental situation.

\section*{Acknowledgements}

We would like to thank Guy Decaudain, David Lloyd Owen, Gyan Bhanot
and   Roch Bourbonnais of TMC for their support.
We also thank H. T. Diep, A. P. Young and D. P. Belanger for various
informations about this subject.

\end{document}